\documentclass[aps,10pt,prl,twocolumn,showpacs,groupedaddress]{revtex4}  % for review and submission
\usepackage{graphicx}  % needed for figures
\usepackage{dcolumn}   % needed for some tables
\usepackage{bm}        % for math
\usepackage{amssymb}   % for math
\usepackage{amsmath}
\begin{document}

\title{Two-particle quantum transmission}
\author{Ron Folman}
	\email{folman@bgu.ac.il}

	\affiliation{Ben-Gurion University of the Negev\\
email: folman@bgu.ac.il, tel: +972-528-795-761, fax: +972-8-6479264}
\date{\today}

\begin{abstract}
Two-photon interference is a fundamental phenomenon in quantum mechanics and stands at the base of numerous experimental observations. Here another manifestation of this phenomenon is described, taking place at a Y junction. Specifically it is shown how the $r^2+t^2$ term which is behind previous observations of two-photon interference, may give rise to different states at a beam-splitter and different two-particle transmission coefficients at a Y junction. Different from previous descriptions of quantum transmission based on one-particle physics, the enhanced transmission described here is due to two-particle physics.
\end{abstract}

\maketitle

Quantum reflection or transmission is a term used when classical probabilities cannot explain a reflection, e.g. when a wave packet hits a potential well, or transmission, e.g. when a wave packet tunnels through a barrier. Numerous effects may be termed quantum reflection, from atoms reflected from a surface \cite{ketterle}, to coherent back scattering of light \cite{nice}, and similarly there are abundant examples of quantum transmission such as electron transport through a ring \cite{joe}. Quantum transmission with larger than classical probabilities is also named quantum tunneling and is the base for numerous physical phenomena (e.g. radioactive decay, cold emission, quantum conductivity) as well as many applications (e.g. scanning tunneling microscope, tunneling diode). These effects are due to the quantum behavior of single particles. In this letter we discuss enhanced transmission due to quantum interference of two particles, specifically photons.

Two-photon interference is one of the most important representations of quantum optics. It has been reviewed in many text books as well as numerous scientific papers. At the base of the effect stands the measured joint probability of the detection of two particles (typically photons) by two detectors. This is usually referred to as intensity correlations and is very different from the amplitude correlations (or first order coherence) which is measured in a typical interference experiment such as the well known double slit experiment.

Two-photon interference has thus far presented itself in several ways including the HBT (Hanbury Brown and Twiss) effect \cite{HBT1954,HBT1956a,HBT1956b,HBT1957,HBT1958,Fano1961,Baym1969,HBT1974} and the HOM (Hong-Ou-Mandel) effect \cite{Hong1987,Ou1989}. Quantitatively, this phenomenon is described by the second order coherence function \cite{Loudon}. This function is defined as
\begin{equation}
g^{(2)}=\langle I_1I_2\rangle /\langle I_1\rangle \langle I_2\rangle
\label{eq:gtwo}
\end{equation}
where $I_1$ and $I_2$ are the currents in the two photo-detectors, and the angled brackets denote the mean. An extensive description of the phenomenon appears in the book of Scully and Zubairy \cite{Scully} and in numerous papers (e.g. Mandel's overview \cite{Mandel}).  The results of a $g^{(2)}$ experiment typically depend on the type of source (e.g. thermal - two independent atoms, correlated photons from a down converter, etc.) and on the phase the particles accumulate on the way to the detectors by virtue of optical elements such as beam splitters or simply distance. For example, in the case of the HBT effect the phases are accumulated due to different paths while in the case of the HOM effect they are accumulated due to an interaction with an optical element. These are technical details which should not mask the common fundamental origin.

\begin{figure}[b]%
\includegraphics[width=\columnwidth]{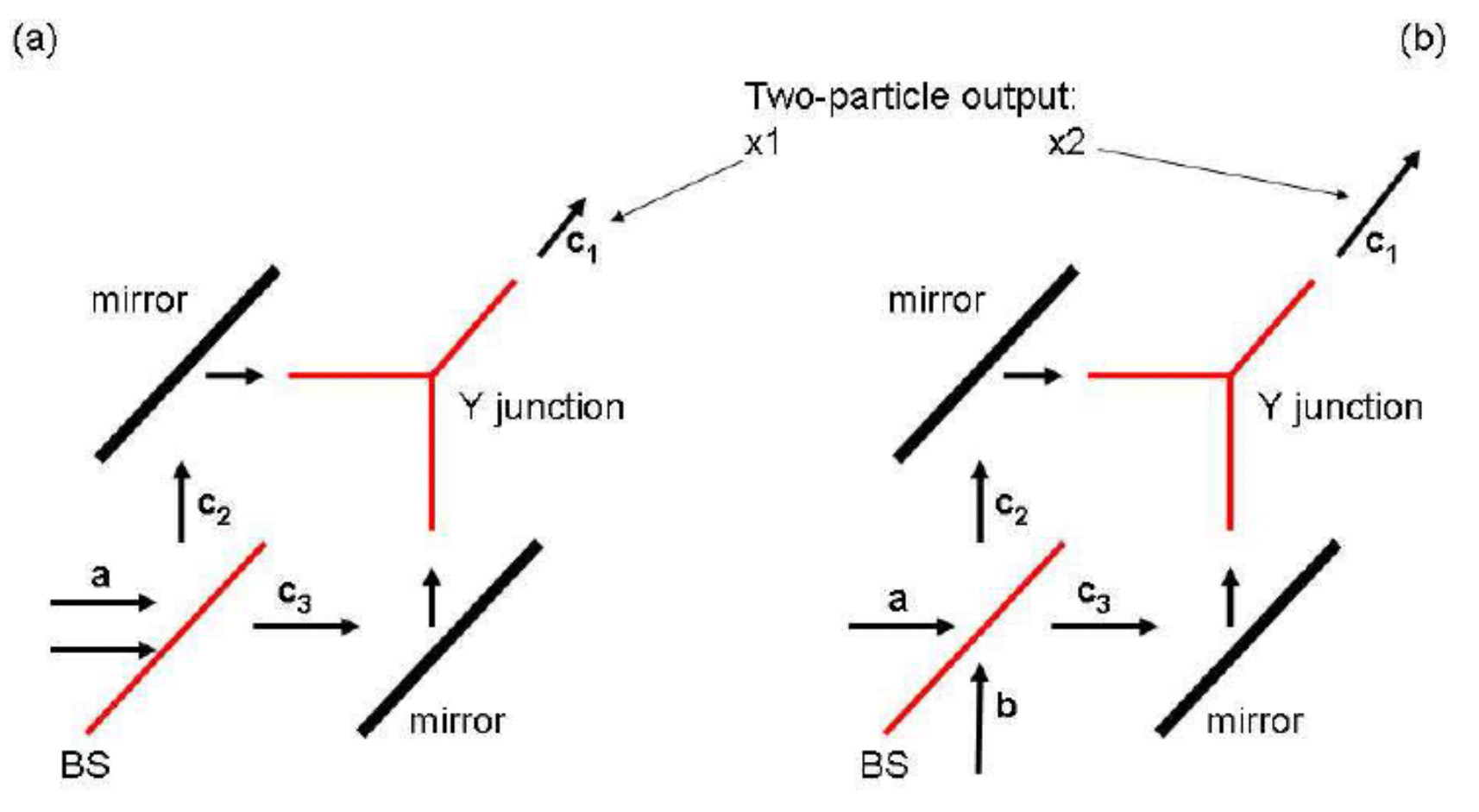}%
\caption{(color online)
Two-photon interference in a Y junction: (a) a PCC state impinging on a Y junction. (b) a HOM state impinging on a Y junction. In the second, the outgoing flux is twice as large as in the first. As the PCC state gives the same flux as classical particles, the HOM state exhibits two-particle quantum transmission.
}%
\label{2types}%
\end{figure}

The beauty of two-photon interference as well as the source of many of the resulting misconceptions lies in the fact that it cannot be viewed as the interference of two waves, as is done in the double slit experiment. This was nicely exhibited by the experiment of Pittman et al. in which the two photon wave packets hit a beam splitter at different times, and still interference was observed \cite{Pittman1996}. Two-photon interference is the interference between two possible occurrences (or histories) which are indistinguishable as they lead to the same observed result.

Photon counting experiments are experiments in which the emission temporal statistics of a source are measured \cite{MandelWolf1970}. These experiments are related to the above experiments in the fact that both are sensitive to the source characteristics, and in the fact that what is measured is the $g^{(2)}$ intensity-intensity correlation as before (here, as a function of time between two photon detections). The difference between spatial coherence and temporal coherence is nicely described in Ref. \cite{Loudon}. In the following I use the term "photon counting experiment" (PCE) or "photon counting configuration" (PCC) simply to note a geometry in which both photons impinge on the beam splitter (BS) from the same port.

In this paper I wish to emphasize a fundamental difference between the PCC and HOM states and show that when they impinge on a Y junction, one acquires a transmission probability which is identical to a classical state while the other acquires a probability twice as large. This effect is due to two-particle interference and may therefore be named two-particle quantum transmission.

PCEs may actually be done by one high temporal resolution detector. This arrangement indeed tests the characteristics of the source which give $g^{(2)}(\tau=0)> 1$ for a chaotic source, $=1$ for a coherent source and $< 1$ for a non-classical source. However, in practical experiments, a $50/50$ BS is added and two detectors are positioned symmetrically relative to this BS so that their individual inability to detect small time gaps between photons, does not inhibit the joint detection of such gaps. Hence, one detector starts the clock while the other stops it.

By analogy, the question may be asked: if a BS introduces such strong detection correlations in the HOM interference experiment, what are the produced correlations when the two photons impinge from the same port?

The usual theoretical treatment suggests that a PCE with one detector is exactly the same as two detectors with a BS. Let us see why. We first define the relation between the incoming and outgoing modes of the BS (Fig. \ref{2types}). If $\hat{a}^\dag$ and $\hat{a}$ are the creation and annihilation operators of the incoming light mode via which the photons impinge on the BS, and if $\hat{c}_k$ and $\hat{c}_k^\dag$ are the operators of the two outgoing modes ($k=2,3$), we may take the relations to be
$\hat{a}^\dag=(\hat{c}_3^\dag+\hat{c}_2^\dag)/\sqrt{2}$ and $\hat{a}=(\hat{c}_3+\hat{c}_2)/\sqrt{2}$. This means that the average photon number (or expectation value) in detector $3$ (in the presence of a BS) is:
\begin{eqnarray}
\bar{n}_3=\langle n|\hat{c}_3^\dag\hat{c}_3|n\rangle =\langle 0|(\hat{a})^n\hat{c}_3^\dag\hat{c}_3(\hat{a}^\dag)^n|0\rangle /n!=\frac12 n
\label{eq:appendixD4}
\end{eqnarray}
and similarly for $\bar{n}_2$. The two-detector correlation is $\langle n_3n_2\rangle =\langle n|\hat{c}_3^\dag\hat{c}_2^\dag\hat{c}_2\hat{c}_3|n\rangle $. Expanding as we did previously this gives $\langle n_3n_2\rangle =\frac{1}{4}n(n-1)$. One thus finds
\begin{equation}
g^{(2)}(\tau)=\frac{\langle n_3n_2\rangle }{\bar{n}_3\bar{n}_2}=\frac{n}{4}(n-1) / \frac{1}{4}n^2=1-\frac{1}{n}
\label{eq:appendixD5}
\end{equation}
which is exactly what one gets without a BS, namely,
\begin{equation}
g^{(2)}(\tau)=\frac{\langle \hat{a}^\dag\hat{a}^\dag\hat{a}\hat{a}\rangle }{\langle \hat{a}^\dag\hat{a}\rangle ^2}.
\label{eq:appendixD6}
\end{equation}

Introducing a phase into the above relations between $\hat{a}$ and $\hat{c}$ changes nothing.

We thus see that indeed the splitting or the phase makes no difference. It is therefore understandable that an analysis of interference effects caused by this BS has to the best of my knowledge been neglected, and that PCEs have been to a large extent described as lying outside the family of two-photon spatial interference experiments. In this letter we show that also PCC states give rise to two-photon interference, and that the outcome of this interference is different than that of the HOM state.

To complete our rather lengthy review, let us now introduce for the sake of unitarity a $\pi/2$ phase into the reflected mode of the BS by defining
$\hat{a}^\dag=(\hat{c}_3^\dag+i\hat{c}_2^\dag)/\sqrt{2}$ and
$\hat{a}=(\hat{c}_3-i\hat{c}_2)/\sqrt{2}$,
and verify that this indeed gives the HOM effect, where the phase does matter.
For the other incoming port we have $\hat{b}^\dag=(i\hat{c}_3^\dag+\hat{c}_2^\dag)/\sqrt{2}$.

For the HOM scenario in which two correlated photons hit the BS, each photon from a different side, we have:
\begin{eqnarray}
\hat{a}^\dag\hat{b}^\dag|0\rangle =\frac12(\hat{c}_3^\dag+i\hat{c}_2^\dag)(i\hat{c}_3^\dag+\hat{c}_2^\dag)|0\rangle=
\frac{i}{\sqrt{2}}(|2,0\rangle +|0,2\rangle )
\label{eq:appendixD8}
\end{eqnarray}
as for bosons the two outgoing operators commute. $|2,0\rangle$ and $|0,2\rangle$ equal $\frac{1}{\sqrt{2}}\hat{c}_{3}^\dag\hat{c}_{3}^\dag|0\rangle$ and $\frac{1}{\sqrt{2}}\hat{c}_{2}^\dag\hat{c}_{2}^\dag|0\rangle$ respectively.

For arbitrary reflection coefficients, the same result may be described in the following way (e.g. \cite{BS}):
\begin{eqnarray}
\sqrt{2}rt|2,0\rangle +\sqrt{2}rt|0,2\rangle +(r^2+t^2)|1,1\rangle
\label{eq:appendixD9}
\end{eqnarray}
where $r$ and $t$ are simply the reflection and transmission amplitudes, and $|1,1\rangle =\hat{c}_{3}^\dag\hat{c}_{2}^\dag|0\rangle$.

\begin{figure}[b]%
\includegraphics[width=\columnwidth]{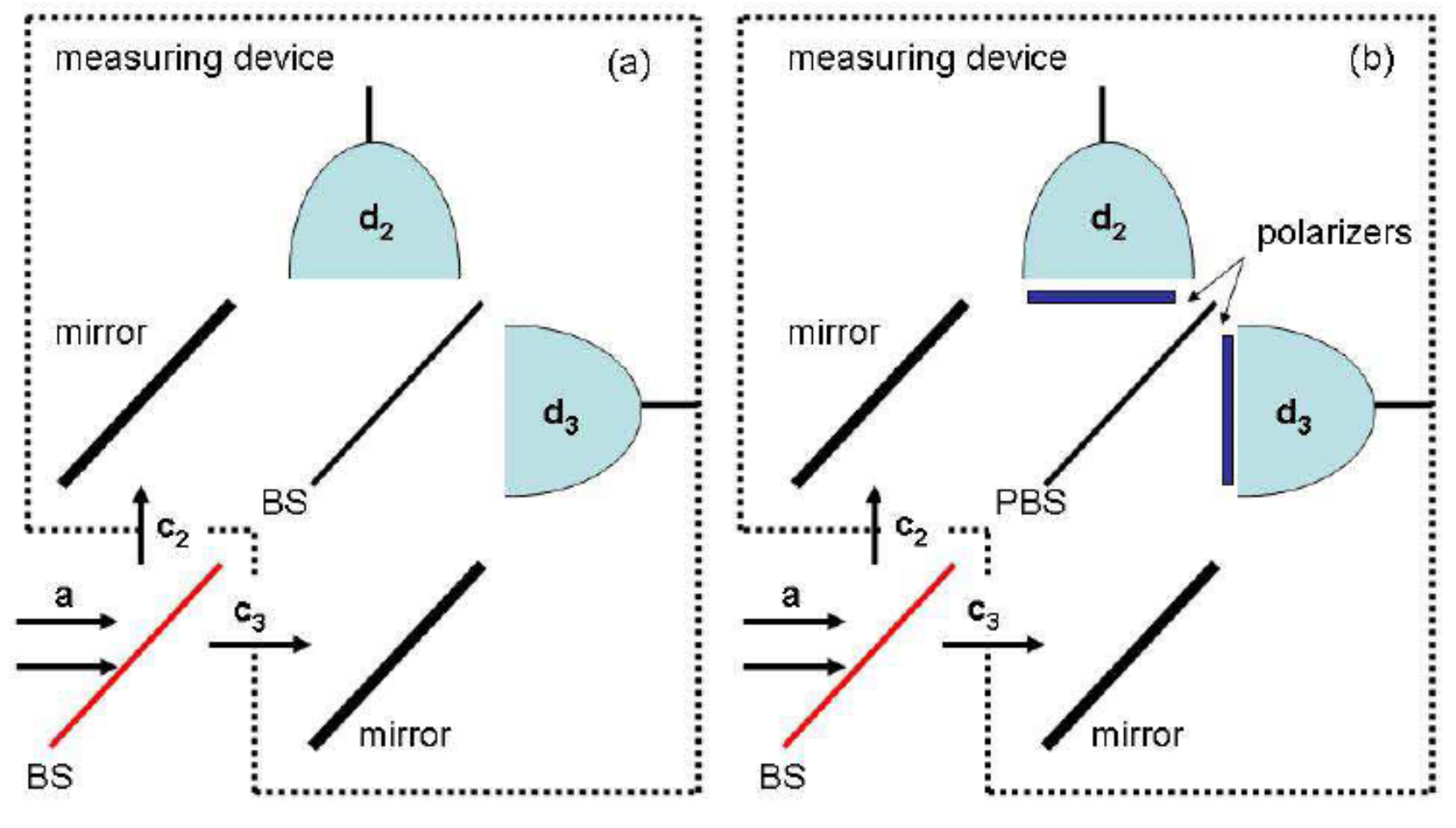}%
\caption{(color online)
measuring two-photon interference in a PCC experiment (beam splitter in red): similar to the HOM state, the $t^2$ and $r^2$ amplitudes are associated with two indistinguishable events. Contrary to the HOM states, the indistinguishable states of the PCC are spatially separated. (a) measured by an external beam splitter and two photo-diodes $d_2$ and $d_3$, and (b) measured by an external polarizing beam splitter with polarizers set to $+45^o$ in front of the detectors.
}%
\label{PhotonCounting}%
\end{figure}

For the case of a $50/50$ BS one has $r=te^{i\frac{\pi}{2}}$. As usual, the squared moduli of the coefficients gives the associated probabilities. We see that the fundamental reason for the HOM two-photon interference resulting in no-coincidence counts lies entirely in the fact that the amplitude of the $|1,1\rangle$ state is $r^2+t^2$ which, for a $50/50$ BS, is just zero.

One may verify that in the case of a PCC, one has:
\begin{eqnarray}
\frac{1}{\sqrt{2}}\hat{a}^\dag\hat{a}^\dag|0\rangle =\frac{1}{\sqrt{2}}(t^2\hat{c}_{3}^\dag\hat{c}_{3}^\dag +r^2 \hat{c}_{2}^\dag\hat{c}_{2}^\dag)|0\rangle+\sqrt{2} rt(\hat{c}_{3}^\dag\hat{c}_{2}^\dag)|0\rangle=\nonumber\\
r^2|2,0\rangle +\sqrt{2}rt|1,1\rangle  +t^2|0,2\rangle.
\label{eq:appendixD10}
\end{eqnarray}

We now go back to the operation of the BS in a PCC. The main message of this letter is to point out that exactly the same $r^2+t^2$ amplitude may be found in a PCC. However, because $r^2$ and $t^2$ terms in the PCC state are associated with spatially separated states (contrary to the HOM state), it may give rise to very different behavior as shown in the following.

Let us begin by noting that there is a unique difference between the HOM state of Eq. (\ref{eq:appendixD9}) and the PCC state described by Eq. (\ref{eq:appendixD10}): while the probabilities of the two surviving bunched states in the former add to one, the probability of the single state in the latter not connected to the $t^2$ and $r^2$ terms amounts to one half. If we could therefore keep only the anti-bunched state in Eq. (\ref{eq:appendixD10}), we would be in fact forcing half of the PCC flux to be reflected back compared to the HOM flux.

Let us analyze an experiment that will do just that. In Fig. \ref{2types} we connect the two output ports of a BS to the two input ports of a Y junction. Such a Y junction may perhaps be realized by the methods described in \cite{anton}. If indeed each outgoing port of our BS could be coupled into a Y junction to combine the two into one mode in which all three states of Eq. (\ref{eq:appendixD10}) are indistinguishable, we would add the three amplitudes to calculate the flux in the combined channel and find that the flux has dropped to $|\sqrt{2}rt|^2=1/2$. We would thus expect $50\%$ of the photons to be reflected back towards the source relative to the HOM state. While previous realizations of quantum reflection or transmission utilize single particle physics, this difference would be due to two-particle physics.

Let us review this with more detail. We have analyzed Y junctions in \cite{ErikaPRL, OfirPRL}. As an example, I utilize a real $3\times3$ transfer matrix which we have analyzed in  \cite{Ofir} and which transfers the vector of three outgoing amplitudes to a vector of three incoming amplitudes:

\[%
\begin{array}
[c]{cc}%
\left(
\begin{array}
[c]{lll}%
~~~1-2T & \sqrt{2T(1-T)} & \sqrt{2T(1-T)}\\
\sqrt{2T(1-T)} & ~~-(1-T) &  ~~~~~~~~T\\
\sqrt{2T(1-T)} &  ~~~~~~~~T & ~~-(1-T)
\end{array}
\right),
\end{array}
\]
where T is the "cross talk" (transfer amplitude) between the two incoming channels. We label the outgoing (combined) port with index $1$ and the two incoming ports (joining BS with Y) as $2$ and $3$. We thus see that $\hat{c}_{2,in}^\dag=\sqrt{2T(1-T)}\hat{c}_{1,out}^\dag-(1-T)\hat{c}_{2,out}^\dag+T\hat{c}_{3,out}^\dag$ and similarly $\hat{c}_{3,in}^\dag=\sqrt{2T(1-T)}\hat{c}_{1,out}^\dag+T\hat{c}_{2,out}^\dag-(1-T)\hat{c}_{3,out}^\dag$. As our incoming state is [Eq. (\ref{eq:appendixD10})] $\frac{1}{\sqrt{2}}r^2(\hat{c}_{2,in}^\dag)^2+\frac{1}{\sqrt{2}}t^2(\hat{c}_{3,in}^\dag)^2 +\sqrt{2}rt\hat{c}_{2,in}^\dag\hat{c}_{3,in}^\dag$, we find the outgoing state to be
\begin{eqnarray}
\frac{1}{\sqrt{2}}r^2(\sqrt{2T(1-T)}\hat{c}_{1,out}^\dag-(1-T)\hat{c}_{2,out}^\dag+T\hat{c}_{3,out}^\dag)^2+\nonumber\\
\frac{1}{\sqrt{2}}t^2(\sqrt{2T(1-T)}\hat{c}_{1,out}^\dag+T\hat{c}_{2,out}^\dag-(1-T)\hat{c}_{3,out}^\dag)^2+\nonumber\\
\sqrt{2}rt(\sqrt{2T(1-T)}\hat{c}_{1,out}^\dag-(1-T)\hat{c}_{2,out}^\dag+T\hat{c}_{3,out}^\dag)\times\nonumber\\
(\sqrt{2T(1-T)}\hat{c}_{1,out}^\dag+T\hat{c}_{2,out}^\dag-(1-T)\hat{c}_{3,out}^\dag).
\label{eq:appendixD16}
\end{eqnarray}

Thus, the probability for having $\hat{c}_{1,out}^\dag \hat{c}_{1,out}^\dag$ is
\begin{eqnarray}
|\frac{1}{\sqrt{2}}(r^2+t^2)+\sqrt{2}rt|^2|[2T(1-T)]|^2=\frac{1}{2}[2T(1-T)]^2,
\label{eq:appendixD17}
\end{eqnarray}
where we have used for the equality $t=\frac{1}{\sqrt{2}}$ and $r=\frac{i}{\sqrt{2}}$.

For the HOM state [Eq. (\ref{eq:appendixD9})], and following the same path which led to Eq. (\ref{eq:appendixD17}), one finds
\begin{eqnarray}
|2rt+(t^2+r^2)|^2|[2T(1-T)]|^2=[2T(1-T)]^2.
\label{eq:appendixD19}
\end{eqnarray}

One may hypothesize that this indicates that bunched states scatter forward better than anti-bunched states, but this is not so as this difference is independent of the value of the reflection and transmission coefficients.

As for indistinguishable particles $\frac{1}{\sqrt{2}}\hat{c}_{1,out}^\dag \hat{c}_{1,out}^\dag=|2,0,0\rangle$, the final transmission (forward scattering) probability of the PCC state would be $[2T(1-T)]^2$ and for the HOM state $2[2T(1-T)]^2$. As expected, the former is smaller by a factor of two relative to the latter.

For a state of two independent photons one finds:
\begin{eqnarray}
[|r+t|^2|\sqrt{2T(1-T)}|^2]^2=[2T(1-T)]^2.
\label{eq:appendixD20}
\end{eqnarray}

As for two independent photons $\hat{c}_{1,out}^\dag \hat{c}_{1,out}^\dag=|2,0,0\rangle$, the final probability for forward scattering is $[2T(1-T)]^2$. We thus find the forward going flux of the PCC state to be the same as the flux of two independent photons. The flux of the HOM state, being twice as large, may then be named quantum enhancement or amplification.

%%%%

It is insightful to see why the above phenomena does not take place in a normal Mach-Zehnder (MZ) interferometer. Let us analyze a PCC state as in Fig. \ref{PhotonCounting}a.

Inserting the definitions of $\hat{c}_3$ and $\hat{c}_2$ in terms of $\hat{d}_3$ and $\hat{d}_2$ into Eq. (\ref{eq:appendixD10}), we find (not noting the vacuum state)
\begin{eqnarray}
\frac{1}{\sqrt{2}}[(t^2(r\hat{d}_{3}^\dag +t\hat{d}_{2}^\dag)(r\hat{d}_{3}^\dag +t\hat{d}_{2}^\dag)+\nonumber\\
r^2(r\hat{d}_{2}^\dag +t\hat{d}_{3}^\dag)(r\hat{d}_{2}^\dag +t\hat{d}_{3}^\dag))+
2rt(r\hat{d}_{3}^\dag +t\hat{d}_{2}^\dag)(r\hat{d}_{2}^\dag +t\hat{d}_{3}^\dag)]=\nonumber\\
(r^4+t^4)|0,2\rangle ^d+2t^2r^2|2,0\rangle ^d+\sqrt{2}rt(t^2+r^2)|1,1\rangle ^d+\nonumber\\
2t^2r^2|2,0\rangle ^d+2t^2r^2|0,2\rangle ^d+\sqrt{2}rt(r^2+t^2)|1,1\rangle ^d=-|2,0\rangle ^d,
\label{eq:appendixD14}
\end{eqnarray}
where the index $d$ denotes the detector plane.

We thus find that just like in the HOM effect, the $t^2+r^2$ factor originating from the first BS has eliminated the probability for coincidence counts in the third term. Furthermore, the above result is an outcome of a rather dramatic destructive interference between the events in which the photons are bunched after the first BS and the event where they are anti-bunched (first and fifth terms), where the first term again arises from the $t^2+r^2$ factor of the first BS. This eliminates the $|0,2\rangle ^d$ state. As two output ports are available to the photon flux, the above interference will not change the total outgoing flux, only the specific port through which the flux exits. In this case, all photons exit through the $|2,0\rangle ^d$ state, a result which is identical to that obtained from two distinguishable photons traversing a MZ interferometer. Similarly, a HOM state (i.e. one photon from each input port of the MZ) will give the same result as two distinguishable photons.

%%%%

Let us complete the discussion by showing how two-particle indistinguishable events which are associated with the $t^2$ and $r^2$ terms (similar to the HOM situation), and which are spatially separated (contrary to the HOM situation), destructively interfere. Contrary to the MZ scheme presented above, the following result cannot be explained as a sum over single photon evolution. As is shown in Fig. \ref{PhotonCounting}b, we replace the second BS with a polarization BS (PBS). We assume that there is no phase difference induced by the PBS between the reflected and transmitted modes. Furthermore, we utilize a source which gives rise to two photons with perpendicular polarizations (e.g. type II down converter). Last, we place $+45^o$ polarizers in front of the detectors. As the operation of the first BS is polarization independent we find for $|\psi\rangle =\hat{a}_{\parallel}^\dag\hat{a}_{\perp}^\dag|0\rangle$ as in Eq.~(\ref{eq:appendixD10})
\begin{eqnarray}
r^2\hat{c}_{2,\parallel}^\dag\hat{c}_{2,\perp}^\dag +rt(\hat{c}_{3,\parallel}^\dag\hat{c}_{2,\perp}^\dag+\hat{c}_{2,\parallel}^\dag\hat{c}_{3,\perp}^\dag) +t^2\hat{c}_{3,\parallel}^\dag\hat{c}_{3,\perp}^\dag.
\label{eq:appendixD15}
\end{eqnarray}

Following the known features of a PBS, we now make the interchanges $\hat{c}_{3,\parallel}^\dag\rightarrow \hat{d}_{3,\parallel}^\dag$, $\hat{c}_{3,\perp}^\dag\rightarrow \hat{d}_{2,\perp}^\dag$, $\hat{c}_{2,\parallel}^\dag\rightarrow \hat{d}_{2,\parallel}^\dag$, and $\hat{c}_{2,\perp}^\dag\rightarrow \hat{d}_{3,\perp}^\dag$. This gives,
\begin{eqnarray}
r^2\hat{d}_{2,\parallel}^\dag\hat{d}_{3,\perp}^\dag +rt(\hat{d}_{3,\parallel}^\dag\hat{d}_{3,\perp}^\dag+\hat{d}_{2,\parallel}^\dag\hat{d}_{2,\perp}^\dag)\nonumber +t^2\hat{d}_{3,\parallel}^\dag\hat{d}_{2,\perp}^\dag=\nonumber\\
r^2(\hat{d}_{2,+45}^\dag+\hat{d}_{2,-45}^\dag)(\hat{d}_{3,+45}^\dag-\hat{d}_{3,-45}^\dag)+\nonumber\\
rt[(\hat{d}_{3,+45}^\dag+\hat{d}_{3,-45}^\dag)(\hat{d}_{3,+45}^\dag-\hat{d}_{3,-45}^\dag)+\nonumber\\
(\hat{d}_{2,+45}^\dag+\hat{d}_{2,-45}^\dag)(\hat{d}_{2,+45}^\dag-\hat{d}_{2,-45}^\dag)]\nonumber\\ +t^2(\hat{d}_{3,+45}^\dag+\hat{d}_{3,-45}^\dag)(\hat{d}_{2,+45}^\dag-\hat{d}_{2,-45}^\dag).
\label{eq:appendixD16}
\end{eqnarray}
where we have ignored the $1/\sqrt{2}$ factors.

Only the terms multiplied by $r^2$ and $t^2$ may give rise to a coincidence count, and as the detectors are preceded by polarizers set at $+45^o$, only the terms $r^2\hat{d}_{2,+45}^\dag\hat{d}_{3,+45}^\dag$ and $t^2\hat{d}_{3,+45}^\dag\hat{d}_{2,+45}^\dag$ may give rise to a coincidence count. As $[\hat{d}_{2}^\dag,\hat{d}_{3}^\dag]=0$ we may join these two terms to give $(r^2+t^2)\hat{d}_{3,+45}^\dag\hat{d}_{2,+45}^\dag=0$. Thus, the coincidence count rate will be zero ($g^{(2)}=0$). This result is a unique indication of two-photon interference and it cannot be imitated by two single photons as these would give rise to a coincidence count with a probability of $2/16$ ($g^{(2)}=\frac12$). Let us note what is trivial but perhaps insightful, and that is that the projection onto the $45^o$ basis by the polarizers acts as a quantum eraser erasing which-path information \cite{eraser0,eraser1,eraser2,eraser3}. The polarizers have erased the information of whether what transpired was a $|0,2\rangle $ or a $|2,0\rangle $ event, these events being very different from the HOM events.

The claim may be made that as we have used two different photons as input we have not made a true analog of the PCC, but rather an analog of the HOM experiment whereby the two polarizations correspond to the two HOM incoming states. This is not so as the $t^2$ and $r^2$ terms giving rise to the effect in the above calculation correspond to spatially separated output states and not to the HOM output states which are not spatially separated. It is this spatial separation of the indistinguishable events which differentiates the HOM and PCC events.

To conclude, we have shown that the $t^2+r^2$ interference term responsible for the HOM effect may be observed also in the photon counting experiment configuration when utilizing the correct measurement basis. Specifically, we have shown that when different two-photon states impinge on a Y junction, one finds different forward going fluxes. These findings represent two-particle quantum transmission.

I would like to thank my good friends and colleagues Carsten Henkel, Yonathan Japha, Daniel Rohrlich, Chris Westbrook and Yasuhiro Tokura for helpful discussions.

\bibliographystyle{apsrev4-1} %Style of Bibliography: plain / apalike / amsalpha / ... names: abbrvnat, cell, apsrev, ieeetr(numbers), short_style, nar, apsrmp4-1long
%\bibliography{literature}

%Merlin.mbs v4.21 2009-07-09.
%

\end{document}